\documentclass[useAMS,usenatbib,letterpaper]{mn2e}
\def\lesssim{\mathrel{\hbox{\rlap{\hbox{\lower4pt\hbox{$\sim$}}}\hbox{$<$}}}}
\def\gtrsim{\mathrel{\hbox{\rlap{\hbox{\lower4pt\hbox{$\sim$}}}\hbox{$>$}}}}

\def\apj{ApJ}                 
\def\apjl{ApJ}                
\def\aap{A\&A}                
\def\mnras{MNRAS}             
\def\ssr{Space Sci. Rev.}

\usepackage[pdftex]{graphicx}
\usepackage{amsmath}
\usepackage{color}
\makeatletter

\def\compoundrel#1\over#2{\mathpalette\compoundreL{{#1}\over{#2}}}
\def\compoundreL#1#2{\compoundREL#1#2}
\def\compoundREL#1#2\over#3{\mathrel
      {\vcenter{\hbox{$\m@th\buildrel{#1#2}\over{#1#3}$}}}}
\usepackage{subfigure}

\title[Crab flares as minijets]{Crab Nebula gamma-ray flares as relativistic reconnection minijets}
\author[E. Clausen-Brown and M. Lyutikov]{E. Clausen-Brown$^{1}$\thanks{browner@purdue.edu} and M. Lyutikov$^1$\thanks{lyutikov@purdue.edu}\\
$^{1}$Department of Physics, Purdue University, 525 Northwestern Avenue, West Lafayette, IN 47907-2036, USA}
\begin{document}

\date{Accepted/Received}

\pagerange{\pageref{firstpage}--\pageref{lastpage}} \pubyear{0000}

\maketitle

\label{firstpage}

\begin{abstract}
The unusually short durations, high luminosities, and high photon energies of the Crab Nebula gamma-ray flares require  relativistic bulk motion of the emitting plasma. We explain the Crab flares as the result of randomly oriented relativistic ``minijets" originating from reconnection events in a magnetically dominated plasma.  We develop a statistical model of the emission from Doppler boosted reconnection minijets and find analytical expressions for the moments of the resulting nebula light curve (e.g. time average, variance, skewness).  The light curve has a flat power spectrum that transitions at short timescales to a decreasing power-law of index 2. The flux distribution from minijets follows a decreasing power-law of index $\sim 1$, implying the average flux from flares is dominated by bright rare events. The predictions for the flares' statistics can be tested against forthcoming observations.  We find the observed flare spectral energy distributions (SEDs) have several notable features: A hard power-law index of $p\lesssim 1$ for accelerated particles that is expected in various reconnection models, including some evidence of a pile-up near the radiation reaction limit. Also, the photon energy at which the SED peaks is higher than that implied by the synchrotron radiation reaction limit, indicating the flare emission regions' Doppler factors are $\gtrsim$ few.  We conclude that magnetic reconnection can be an important, if not dominant, mechanism of particle acceleration within the nebula.  
\end{abstract}

\section{Introduction}
\label{intro}
The presumed constancy of the high energy Crab nebula emission has surprisingly been shown to be false by multiple day- to week-long flares, presenting a challenge to standard pulsar wind models \citep{Kennel:1984}.  During these events, the Crab Nebula gamma-ray flux above 100 MeV exceeded its average value by a factor of several or higher \citep{Abdo:2011,Tavani:2011,Buehler:2012,Striani:2011}, while in other energy bands nothing unusual was observed \citep[e.g.][and references therein]{Abdo:2011,Tavani:2011}.  Additionally, sub-flare variability timescales of $\sim 10$ hours has been observed \citep{Balbo:2011,Buehler:2012}.

There are two interesting observational facts related to the gamma-ray flares.  First, is their unusually short duration of a few days.  This time scale, on the one hand, is millions of times longer than the period of the pulsar, yet on the other hand it is hundreds of times shorter than the nebula's dynamical timescale of $\sim$ few years.  We consider it unlikely that the flare is related to the changing plasma properties within the pulsar's light cylinder, both due to the extremely large separation of temporal scales and due to the fact that no changes in the radio pulsar timing properties were seen during the flare \citep{Espinoza:2010}.  Thus, we associate the duration of the flare with stochastically changing properties of plasma within the nebula.  Second, the flaring behavior consists of apparently isolated, intermittent events that are dominated by bright rare flares.  


Two contrasting models of these flares can be envisioned. First, a flare can be due to large scale changes in the steady nebular flow, amplified by the effects of relativistic beaming.  Operating within this model, \cite{Komissarov:2011} and \cite{Lyutikov:2011} place the flaring location in the downstream region of an oblique shock.  The post-shock flow of an oblique shock can be highly relativistic, and thus Doppler boosted, with a bulk Lorentz factor of $\Gamma \sim \phi_s^{-1}$, where $\phi_s$ is the angle between the upstream velocity and the shock plane.  The short timescale can be attributed to the shock normal changing direction (perhaps due to sausage or kink instabilities, or a corrugation perturbation), causing the post-shock velocity to sweep across the line of sight and create a short flare in Doppler boosted emission (the lighthouse effect).  

Second, the flare can be due to a highly localized emission region, or blob, so that the flare observables determine the intrinsic properties of the emission region.  The natural flaring mechanism in this category is relativistic magnetic reconnection --- the focus of this work --- which has been invoked by Crab Nebula flare models \citep{Bednarek:2011,Cerutti:2012,Uzdensky:2011b} and fast flaring models in gamma-ray bursts (GRBs) and active galactic nuclei \citep[AGN,][]{Lyutikov:2001,Lyutikov:2003b,Lyutikov:2006a,Giannios:2009,Giannios:2010,McKinney:2010}.  

To see why reconnection is a natural flaring mechanism in PWNe, consider that the energy budget of the pulsar wind is set by the spin-down power, $P_{\rm spin}$, of the pulsar.  In the standard model, $P_{\rm spin}$ is a smooth, monotonically decreasing function of time, implying that the nebula emission will track this smooth decline. In contrast, the emission from relativistic outflows in systems such as GRBs and AGNs are probably related to an irregular accretion process.  Their variable accretion rates and other accretion disk instabilities may produce unsteady outflows with internal shocks, providing a viable alternative to reconnection as a flaring mechanism in the outflow \citep[e.g.][]{Rees:1994,Spada:2001}.  In pulsar outflows, however, the internal shock scenario is more difficult to realize given the regular behavior of $P_{\rm spin}$. Instead, it is more likely that day-scale flaring is due to the build up and eventual release of free energy somewhere in the PWN\footnote{The wisps in the nebula that exhibit variability timescales of $\sim$ months, are not due to an unsteady wind, but are probably related to the ion-affected structure of the wind termination shock \citep{Gallant:1994} or a synchrotron instability \citep{Hester:2002}.  However, the recent discovery of the several year $\sim10$\% decrease in the nebular X-ray emission \citep{Wilson-Hodge:2011} is puzzling in light of the Crab's smooth spin-down rate, but the unknown mechanism for this emission decrease is probably unrelated to day-long gamma-ray flares, since its timescale is several orders of magnitude longer.}.  Magnetic reconnection is a natural candidate for this process as it is the primary mechanism by which magnetic free energy may be suddenly converted into particle kinetic energy, which occurs via the topological rearrangement of magnetic field lines.  Furthermore, in the closest space laboratory available, the solar system, this is precisely what is observed, where intermittent flaring and the production of non-thermal particles is regularly associated with magnetic reconnection \citep{Priest:2000}.

In this work we locate the reconnecting plasma at multiple sites in the nebula, and presume the reconnection plasma's Alfv\'en velocity approaches the speed of light.  The reconnection outflow speeds are then relativistic \citep{Lyutikov:2003c,Lyubarsky:2005} and behave as ``minijets," a model that has been used to overcome the gamma-ray opacity problem in the context of gamma-ray bursts and active galactic nuclei \citep{Lyutikov:2003b,Lyutikov:2006a,Giannios:2009}.  In an approach similar to ours, \cite{Yuan:2011} invoke a series of relativistically moving ``knots" (``minijets" in our terminology), with statistical properties that are compared to the Crab Nebula light curve via Monte Carlo simulations (they do not identify the knots with reconnection).  Our model differs from that of \cite{Yuan:2011} in that it is analytical, and it presumes that relativistic beaming controls the observed statistical properties of the reconnection minijets.

This paper is organized as follows.  In \S\ref{estimates} we first focus on constraints to the emission regions' magnetic field derived from adiabatic expansion and synchrotron cooling, and compare these to Crab Nebula magnetic field estimates based on equipartition and spectral modeling.  We include the possibility that the emission region has a bulk relativistic velocity with a Lorentz factor of $\Gamma$ and associated Doppler factor of $\delta=(\Gamma-\sqrt{\Gamma^2-1}\cos{\theta_{ob}})^{-1}$, where $\theta_{ob}$ is the angle between the blob velocity and the line of sight.  Due to the high luminosity and short variability of the flare, we assume $\theta_{ob}\approx 0$, in which case $\delta \approx 2\Gamma$.  We then speculate in \S\ref{reconnection} that the Crab Nebula flare originates from a relativistic outflow caused by magnetic reconnection in a high-$\sigma$ plasma \citep[$\sigma=$ Poynting flux/particle flux,][]{Kennel:1984}. \S\ref{SED} focuses on the emission region spectral energy distribution (SED) in the context of the synchrotron radiation reaction cut-off and the functional form of the observed SEDs .  In \S\ref{stats}, we construct an analytical statistical model which describes the statistics of both individual flares and the entire light curve's statistical moments (e.g. average and variance) and its power spectrum.  We provide a summary of our conclusions in \S\ref{conclusion}.

\section{Emission region parameter estimates}
\label{estimates}
In this section the magnetic field of the emission region, or blob, is constrained under two separate assumptions: that (i) adiabatic cooling or (ii) synchrotron cooling controls the flare duration.  The relevant observed flare parameters are the flare duration, $\tau=10^{5.5} \tau_{5.5}$ s; typical photon energy, $\epsilon_p=100\epsilon_{100}$ MeV; the blob's isotropic equivalent luminosity, $L_{iso}=10^{36.6}L_{36.6}$ ergs s$^{-1}$; and blob magnetic field, $B=0.3B_{\rm eq}$ G; where the fiducial flare parameter values are based on the Fermi/LAT data of the September 2010 flare \citep{Abdo:2011} and the nebula equipartition magnetic field \citep{Trimble:1983}.  

Assuming adiabatic cooling limits the flare duration, a lower limit on the blob frame magnetic field can be made by examining the energy content of the blob.  We parametrize the total magnetic energy of the blob as being some significant fraction $f_M$ of the blob's total energy content, $f_M E'=B'^2R'^3/6$.  Assuming the blob is causally connected and is moving directly along the line of sight, the blob radius is $R'\leq c \Gamma \tau$.  Of the total energy content of the blob, $E= f_{M}^{-1} \Gamma R'^3 B'^2/6$, if a fraction $f_{rad}$ is radiated away during the flare, then the total radiated power is $L=f_{rad} E/\tau$.  $L$ is related to the isotropic equivalent luminosity, $L_{iso}$, by assuming that all of $L$ is beamed into a solid angle of $\pi \Gamma^{-2}$ which leads to the relation, $L_{iso}=4\Gamma^2L$. Thus, the isotropic equivalent luminosity can be expressed as 
\begin{gather}
L_{iso}=\frac{2}{3}f_{rad} f_M^{-1}R'^3 B'^2 \tau^{-1} \Gamma^3.
\label{Liso}
\end{gather}
In order to use equation (\ref{Liso}) to constrain the magnetic field, $R'$ is estimated, as before, using the light crossing time $\tau>R'(\Gamma c)^{-1}$.  This constraint may be interpreted as the blob slow down time \citep[see, e.g.,][]{Giannios:2009} or as the adiabatic cooling time.  That is, if the blob begins with an initial radius $R'$ and expands at near light speed, $R_t'=R'+ct'$ then the blob light crossing time, $R'/c$, is the time the blob takes to double in radius and adiabatically cool by a significant amount.  Solving equation (\ref{Liso}) for $B'$ and using the inequality $R'\leq c \Gamma \tau$ allows us to set a lower limit on $B'$:
\begin{gather}
B' > 1.5 (f_M/f_{rad})^{1/2}L_{36.6}^{1/2} \tau_{5.5}^{-1} \Gamma^{-3} \quad \mbox{mG}.
\label{energetics}
\end{gather}
If we assume 10\% of the blob's energy content is radiated away ($f_{rad}=0.1$) and the blob is magnetically dominated ($f_M\sim1$), then $B'>4.7$ mG.  Typical PWN models suggest the pulsar wind begins magnetically dominated, and somehow transitions to being cold and particle dominated near the termination shock so that the magnetization parameter is $\sigma=B'^2(8\pi\rho' c^2)^{-1}=3\times 10^{-3}$, where $\rho$ is the mass density \citep{Kennel:1984}.  However, exactly how $\sigma$ evolves from $\sigma\gg1$ near the pulsar to $\sigma \ll 1$ at the termination shock is difficult to explain, provoking competing models that suggest $\sigma \gtrsim 1$ at the shock \citep{Begelman:1998,Lyutikov:2010}.  Thus, near the inferred location of the reverse shock the blob could be magnetically dominated so that $f_M\sim 1$ (in a cold plasma, $f_M=\sigma/(1+\sigma)$) and $B'\gg$ the standard nebula value (0.3 mG), or in the canonical model where the plasma is not magnetically dominated, $f_M$ may be small and $B'$ within the measured range of $\sim 0.1$ mG.

The synchrotron radiative cooling timescale sets a more robust lower limit on $B'$ because it avoids the uncertainty in $f_r$ and $f_M$ that appear in equation (\ref{energetics}).  The synchrotron cooling time, $\tau_c= 5 \cdot 10^8 (2\Gamma \gamma B'^2)^{-1}$ s, can be found given the typical synchrotron photon energy of $\epsilon_p=2\Gamma \gamma^2 h e B'/(m_e c)$, where $\gamma$ is the lepton Lorentz factor, $h$ is Planck's constant, $e$ is the elementary charge, and $m_e$ is the lepton mass. These two expressions combined give a synchrotron cooling time of
\begin{gather}
\tau_c= 21 \epsilon_{100}^{-1/2} B_{\rm eq}'^{-3/2} \Gamma^{-1/2} \quad \mbox{days}.
\end{gather}
If a significant fraction of the blob energy is dissipated via synchrotron radiation in the flare, then $\tau \geq \tau_c$ and a lower limit can be set on the blob frame magnetic field:
\begin{gather}
B'> 0.97 \epsilon_{100}^{-1/3}\tau_{5.5}^{-2/3}\Gamma^{-1/3} \quad \mbox{mG}.
\label{cooling}
\end{gather}
Note that this lower limit \cite[which is similar to that found in][]{Bednarek:2011} exceeds the observed values of the nebular magnetic field ($0.1$ to $0.3$ mG), unless relativistic motion is invoked.

The upshot of these parameter estimates is that the magnetic field of the emission region appears to be significantly higher than both nebula equipartition field estimates \citep[0.3 mG,][]{Trimble:1983} and measurements based on modeling of the nebula SED \citep[0.1 mG,][]{Abdo:2010a}.  This suggests that the flaring either occurred in a region of high magnetic field in the nebula, or that the emission region is moving toward Earth with $\Gamma\gtrsim$ few, which reduces the above magnetic field estimates. 

\section{Magnetic reconnection?}
\label{reconnection}
Magnetic reconnection provides a natural explanation of the implied relativistic motion discussed above, the intrinsic short timescales, and the flares' intermittency. Reconnection is a process in which the magnetic energy of a localized region, a current sheet, is suddenly converted to random particle energy, and bulk relativistic motion \citep[for studies on relativistic reconnection, see][]{Blackman:1994,Lyutikov:2003c,Lyubarsky:2005,Uzdensky:2011a}.  With regard to PWNe, reconnection has already been studied as a possible resolution of the well known $\sigma$-problem \citep{Coroniti:1990,Lyubarsky:2001}.  In the canonical PWN model \citep{Rees:1974,Kennel:1984}, the magnetization parameter $\sigma$ is high only for radii that are well within the wind termination shock, $r_s$, and reduces to $\sigma \sim 10^{-3}$ to $10^{-2}$ near $r_s$.  To explain how $\sigma$ is reduced so drastically as the plasma propagates out to $r_s$, reconnection in a striped wind has been invoked as the mechanism by which magnetic energy is transferred to particle energy, thereby reducing $\sigma$.  In a challenge to the canonical PWN model, \cite{Begelman:1998} argues that the toroidally dominated large-scale nebular magnetic field is subject to the $m=1$ kink mode instability near the inferred location of $r_s$, causing the nebular field to have coherence lengths on the order of $r_s$ instead of the size of the radio nebula as it is in canonical models.  This obviates the need for such low $\sigma$ at $r_s$ and may cause reconnection throughout the nebula.  In a similar vein, \cite{Lyutikov:2010} proposes a model in which reconnection occurs primarily along the rotation axis and equatorial region well beyond the light cylinder, thus qualitatively reproducing the jet/equatorial wisp morphology of the nebula. 

In our simple model, we propose that multiple reconnection sites are located in a region of the inner nebula with a magnetization parameter of order $\sigma\sim$ few, thereby accelerating mildly relativistic outflows \citep{Lyutikov:2003c,Lyubarsky:2005}.  We assume the nebular magnetic field is similar to the reconnection minijet's emission region magnetic field, $B_{\rm neb}\sim B'$.  The observational nebular magnetic field estimates range from $B_{\rm neb}\sim0.14$ to $0.3$ mG \citep{Trimble:1983,Abdo:2010a}.  These values of $B_{\rm neb}$ are clearly less than the lower limit set by the cooling equation (\ref{cooling}) of $B'>1$ mG unless we impose relativistic motion.  Solving for the Lorentz factor in equation (\ref{cooling}), we find
\begin{gather}
\Gamma>34E_{100}t_{5.5}^{-2} B_{\rm eq}^{-3}.
\label{coolGamma}
\end{gather}
Such a high Lorentz factor may not be required if the magnetic field is higher near the reconnection region.  In split monopole models of PWNe, the toroidally dominated magnetic field ($B_{\phi}\propto \sin{\theta}$, where $\theta$ is the polar angle) is highest in the equatorial regions \citep{Michel:1973,Bogovalov:1999}.  Thus, if reconnection occurs in the equatorial current sheet or equatorial striped wind, then the magnetic field will indeed be higher then the nebula average of $0.14-0.3$ mG.  In MHD simulations of the Crab pulsar winds, some parts of the nebula magnetic field reach $0.45$ to $0.6$ mG \citep{Komissarov:2004,Komissarov:2011} which requires that the Lorentz factor be at least 4 to 10 according equation (\ref{coolGamma}).  We note that it is also possible there is no relativistic motion, and the flaring region simply has a high magnetic field of $\gtrsim 1$ mG, a value that is indeed found in bright localized regions of the nebula \citep[e.g. the inner ``wisps" and ``knots" of the nebula,][]{Hester:1995}.

\section{Minijet SED}
\label{SED}
The spectral energy distribution (SED) produced in a flaring region may be significantly influenced by the proximity of the emitting particles to the synchrotron radiation reaction limit implied by MHD such that the electron distribution is either mono-energetic or a power-law with an abrupt cut-off.  If a mono-energetic electron distribution is not formed and magnetic reconnection is what causes the flares, then reconnection may leave an imprint on the power-law index of the accelerated particles in the form of a hard electron distribution.  Here we discuss these features in relation to the observed SEDs from the April 2011 flare and suggest these observations point toward relativistic reconnection minijets.

The best Crab Nebula flare SED observations to date are 11 SEDs taken during the April 2011 flare by the Fermi/LAT team \citep{Buehler:2012}.  \cite{Buehler:2012} fit the SEDs with an empirical function of the form 
\begin{align}
N_{\epsilon}=A\epsilon^{-\gamma_F}\exp{(-\epsilon/\epsilon_c)}, 
\label{fit}
\end{align}
where $\epsilon$ represents photon energy, and different values for $A$ and $\epsilon_c$ were used for each SED fit.  The parameter $\gamma_F$ was assumed to be constant for all of the SEDs, and its best fit was found to be $\gamma_F=1.27 \pm 0.12$.  Significantly, the observed integrated flux above 100 MeV, $F$, correlates with the flare SEDs peak energy, $\epsilon_{\rm peak}$, such that $d \log{F}/d \log{\epsilon_{\rm peak}}=3.42\pm0.85$ as expected in simple models of Doppler beaming \citep{Lind:1985}.   In the context of the above observations, we discuss the development of an effectively mono-energetic electron distribution near the radiation reaction limit from a hard electron distribution of the type expected in magnetic reconnection.

Mono-energetic distributions have already been examined in the context of the Crab flares in \cite{Uzdensky:2011b} and \cite{Cerutti:2012}, in which a specific model of reconnection causes electrostatic particle acceleration that produces an approximate mono-energetic electron distribution near the energy associated with the electrostatic potential drop.  Unlike the model we discuss here, the \cite{Uzdensky:2011b} model avoids the MHD limit on particle energy and displays no bulk relativistic motion by invoking a specific geometry of the reconnection region.  In this section, we do not make such geometrical requirements, instead we focus on a general discussion of how the radiation reaction limit can affect flare SEDs.

\begin{figure}
\centering
\includegraphics[width=3.35in]{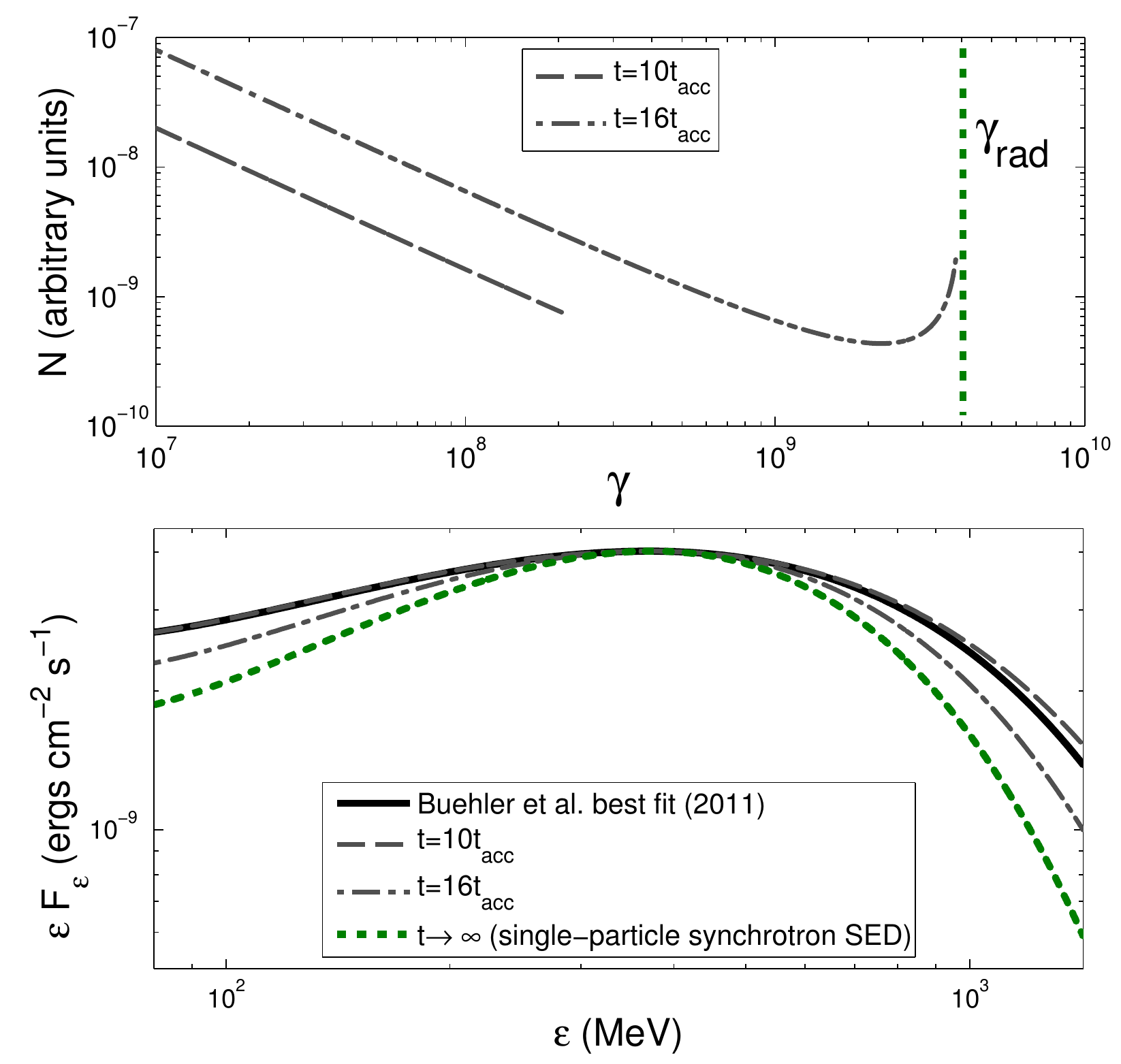}
\caption{\textit{Top plot.}  A solution for $N(\gamma,t)$ (see eqn. \ref{N} in text) is shown for $t=10t_{acc}$ and $16t_{acc}$, with different normalizations to ease visual comparison.  The SEDs corresponding to $N(\gamma,t)$ at $t=10t_{acc}$ and $16t_{acc}$ are in the bottom plot.  For $t=10t_{acc}$, the solution is a pure power-law of $N\propto \gamma^{-1.1}$; for $t=16t_{acc}$, a pile-up develops short of $\gamma=\gamma_{rad}$.  \textit{Bottom plot.} Plotted here are four different flare SEDs added onto the average nebula SED: equation (\ref{fit}) with the parameters reported in \protect\cite{Buehler:2012}, an SED derived from a power-law electron distribution of $N=K_e\gamma^{-1.1}d\gamma$ and $N=0$ for $\gamma>\gamma_{\rm max}$ (i.e. the $t=10t_{acc}$ solution to eqn \ref{N}), the SED derived from the pile-up solution to equation (\ref{N}) for $t=16t_{acc}$, and a single-particle synchrotron SED.  For visual comparison, all three SEDs are adjusted so they have the same maximum and peak energy, otherwise the $t=10t_{acc}$ solution would peak at a lower energy than the solutions for $t=16t_{acc}$ and $t \rightarrow \infty$ (the mono-energetic solution).}
\label{NSED} 
\end{figure}

A critical feature in the SED of the Crab Nebula flares that has already received attention is the synchrotron emission that is above the classical synchrotron limit of $\sim 10^2$ MeV \citep{deJager:1996,Lyutikov:2010,Abdo:2011}.  Importantly, the electron distribution function may display a power-law with an excess, or pile-up, of particles near the synchrotron limit, in which case the emitting particles will display a SED that is close to the single-particle synchrotron SED.  This limit comes from ideal MHD, in which the electric field, $E$, is less than the magnetic field such that $E=\eta B$, where $0<\eta<1$.  To illustrate these points, we briefly examine the particle acceleration process.  First, we assume a particle's energy is approximately described by
\begin{align}
\frac{d\gamma}{dt}&\approx \frac{eE}{m_ec}-\beta_s\gamma^2 \notag \\
&=\eta \omega_B-\beta_s\gamma^2
\label{acc}
\end{align}
where $\beta_s=2/3e^4B_{\perp}^2/(m_e^3c^5)$, $E=\eta B$, and $\omega_B=eB/m_ec$.  The first term describes particle acceleration by the electric field (where $E$ is only an approximation of the true electric field since the accelerating particle velocity is not always parallel to the electric field) and the second term describes synchrotron losses.  We expect the initial particle population to be accelerated to higher energies until $d\gamma/dt=0$, where the synchrotron energy losses equal the acceleration rate such that $\eta \omega_B=\beta_s\gamma^2$.  If ideal MHD holds in the acceleration region, then $\eta<1$, which in turn implies the existence of a maximum possible Lorentz factor allowed by the synchrotron radiation backreaction for a given magnetic field value:
\begin{align}
\gamma_{\rm rad}=\sqrt{\frac{3m_e^2c^4\eta}{2e^3B_{\perp}}}=5\times 10^9 \eta^{1/2}B_{\perp,\rm eq}^{-1/2}.
\label{gamma_rad}
\end{align}

To quantify what we mean by an electron distribution function that ``piles-up" near $\gamma_{\rm rad}$ \citep[e.g.][]{Schlickeiser:1984}, we examine time evolution of the electron distribution function $N(t,\gamma)d\gamma$, which describes the number of electrons with a Lorentz factor between $\gamma$ and $\gamma+d\gamma$ at time $t$.  The time evolution of $N(\gamma,t)$ in our toy model is described by the equation \citep{Kirk:1998}
\begin{align}
\frac{\partial N}{\partial t}+\frac{\partial}{\partial\gamma} \left[\left(\frac{\gamma}{t_{acc}}-\beta_s\gamma^2\right)N\right]+\frac{N}{t_{esc}}=Q\tilde{\delta}(\gamma-\gamma_0),
\label{N}
\end{align}
where the acceleration time $t_{acc}$ and the escape time $t_{esc}$ are constants, $N(\gamma,0)=0$, $Q$ is the injection rate of particles with Lorentz factors of $\gamma_0$, and $\tilde{\delta}$ is the Dirac-delta.  Equation (\ref{N}) describes the time evolution of non-thermal particles subject to mono-energetic injection, particle acceleration and escape, and synchrotron radiation.  The solution to equation (\ref{N}) subject to the above conditions can be found in \cite{Kirk:1998}.  In this solution, the electron distribution, $N$, becomes mono-energetic as $t \rightarrow \infty$  (i.e. $N\sim \tilde{\delta}(\gamma-\gamma_{\rm rad})$) if the particle injection process ceases and the acceleration process continues, since all the injected electrons will simply be accelerated up to $\gamma_{\rm rad}$.  Even if the injection process continues during the acceleration process, then the distribution can still develop a ``pile-up" just below $\gamma_{\rm rad}$, effectively becoming mono-energetic as discussed below (see also fig. \ref{NSED}).  However, such a pile-up distribution only occurs depending on the details of the injected spectrum of particles and particle escape.  In the particular case described above, a pile-up only occurs if $t_{esc}>t_{acc}$ \citep[see also][]{Schlickeiser:1984}.  

\begin{figure}
\centering
\includegraphics[width=3.35in]{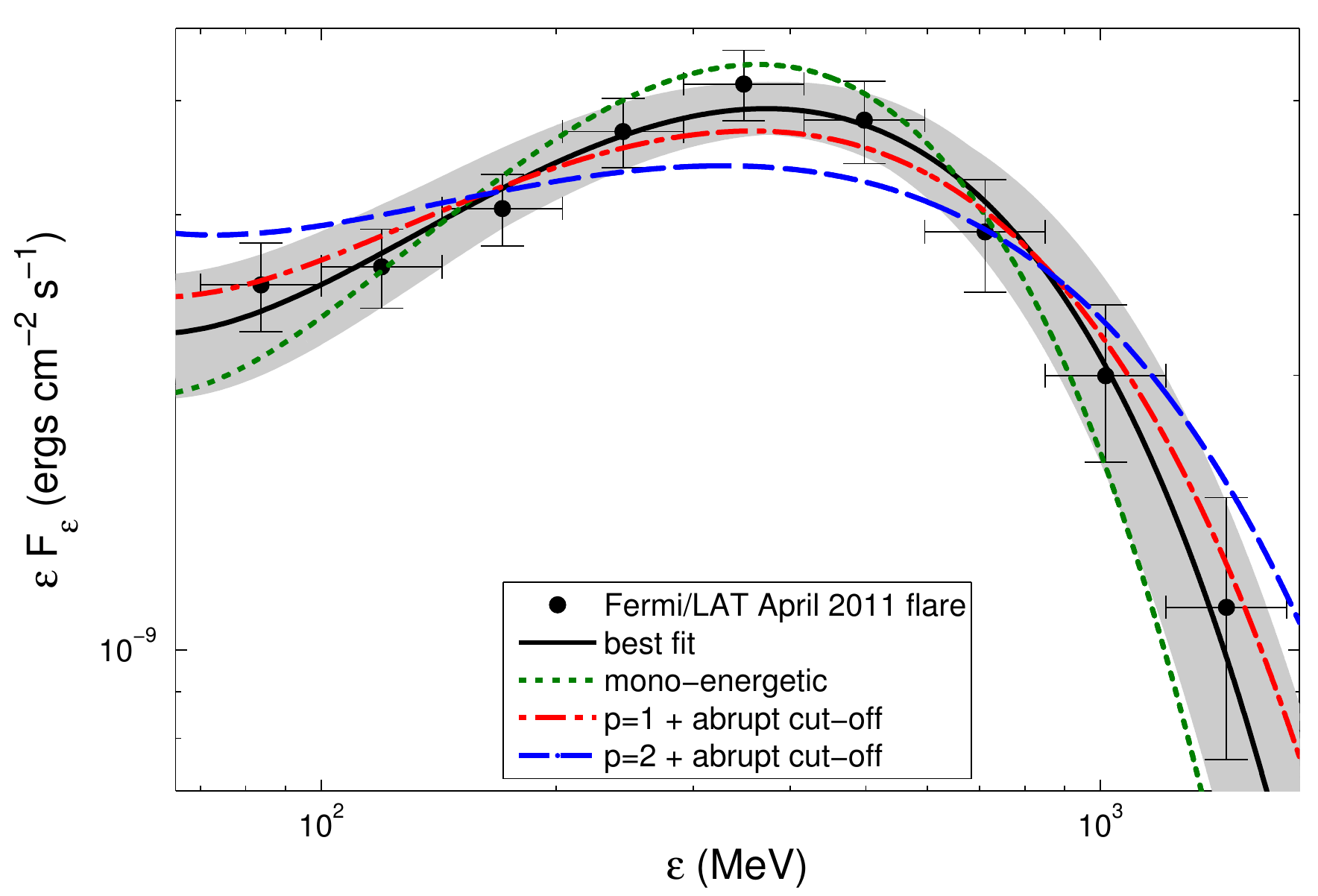}
\caption{The Fermi/LAT data from the most energetic part of the April 2011 flare are shown \protect\citep[figure 6, panel 7 of][]{Buehler:2012} with the corresponding best fit curve from equation (\ref{fit}) and SEDs from three different electron energy distributions: a $p=1$ power-law with an abrupt cut-off, the same for $p=2$, and a mono-energetic electron distribution.  All of the curves are summed with the constant nebula SED component used in \protect\cite{Buehler:2012}.  The shaded area represents the one-$\sigma$ error region (see text for more details).}
\label{bestfit} 
\end{figure}

We have plotted the solution to equation (\ref{N}) at two time-slices, $t=10t_{acc}$ and $t=16t_{acc}$, in figure \ref{NSED} for the parameter choices $t_{esc}=10t_{acc}$ and $\gamma_{\rm rad}=4\times 10^9$.  As figure \ref{NSED} illustrates, a pile-up does eventually accumulate just short of $\gamma=\gamma_{\rm rad}$, eventually evolving to a mono-energetic distribution that produces a single-particle synchrotron SED as shown in the bottom panel.  Visual inspection of figure \ref{NSED} (bottom panel) makes it clear that the power-law distribution of $N\propto \gamma^{-1}$ compares more favorably than the mono-energetic distribution with the best fit of the April 2011 flare SEDs reported in \cite{Buehler:2012}.  

Unlike \cite{Buehler:2012}, we examine only the most luminous observed SED from the April 2011 flare \citep[figure 6, panel 7 in][]{Buehler:2012}, since over the course of the nine-day flare, the functional form of the SED may have significantly changed, and this SED is least affected by the background nebula emission.  As shown in figure \ref{bestfit}, we fit the SED to equation ({\ref{fit}), which yields $\gamma_F=1.08 \pm 0.16$ with a reduced $\chi^2$ of 0.35.  The shaded area in figure (\ref{bestfit}) is the ``one-$\sigma$ error region."  It represents a subset of curves generated by equation (\ref{fit}), each of whose parameters lie within a one-$\sigma$ interval\footnote{The error analysis was done via Monte Carlo simulation.  We assume Gaussian errors in the measured $\epsilon F_{\epsilon}$ values, then randomly generate $10^4$ synthetic SEDs, and fit each synthetic SED to equation (\ref{fit}).  The resulting spread about the best fit value for each parameter in equation (\ref{fit}) allowed the calculation of each parameter's standard deviation, or $\sigma$.} of the best fit parameters.
  
Equation (\ref{fit}) is a useful SED fitting function for the relevant energy range because it has the same functional form (to within a few percent) as SEDs derived from three different electron distributions: power-laws distributions with abrupt cut-offs, power-law distributions with pile-ups (as in fig. \ref{NSED}), and mono-energetic distributions.  Note, however, that the functional form of SEDs derived from power-law distributions that are near the $\epsilon F_{\epsilon}$ peak depend sensitively on the form of the electron distribution cut-off.  For exponential cut-offs to the electron distribution function, the derived SED is broader than SEDs derived from power-laws with abrupt cut-offs that are approximated by equation (\ref{fit}).  Here we presume the radiation reaction limit is the mechanism whereby either an abrupt cut-off or a pile-up develops in the electron energy distribution.  Also, near the $\epsilon F_{\epsilon}$ peak, the $\gamma_F$ parameter cannot be interpreted in the usual way as being equal to $(p+1)/2$ for power-law distributions, or $2/3$ for mono-energetic distributions.  For example, SEDs corresponding to $p=1$ and $p=2$ are well approximated by $\gamma_F\approx 1.30$ and $1.56$ respectively, while a mono-energetic distribution has $\gamma_F\approx0.7$ \citep[e.g.][]{Melrose:1980}; for the purpose of comparison, the best fits of these distributions (which only vary normalization and cut-off energy since $p$ is fixed) are plotted in figure \ref{bestfit} next to the best fit of equation (\ref{fit}).   

The best fit value of $\gamma_F=1.08 \pm 0.16$ lies in between that expected from a mono-energetic distribution and the $p=1$ distribution, while distributions for which $p \geq 2$ are unlikely to have produced the observed SED. The precise $p$ corresponding to the best fit of $\gamma_F=1.08 \pm 0.16$ is $p\sim -0.2^{+0.9}_{-2}$, a large range of values due to the calculated SED's weak dependence on $p$ for $p\lesssim 0$ for the energy range in question.  The best fit values for $\gamma_F$ may imply the development of a pile-up near $\gamma_{\rm rad}$ from a hard power-law electron distribution of $p\sim 1$ as seen in figure \ref{NSED} and discussed in, for example, \cite{Schlickeiser:1984}.  A pile-up pushes a $p=1$ SED's $\gamma_F$ value down from $\gamma_F\approx 1.30$ to that of the mono-energetic distribution, $\gamma_F=0.7$, as illustrated in figure \ref{NSED}.  We verify that pile-up scenario could explain the data through the toy model we develop from equation (\ref{N}) by setting $\gamma_0=10^4$, $\gamma_{\rm rad}=4\times 10^9$, and $t_{\rm esc} \gg t_{\rm acc}$, after which we find the best fit $\gamma_F$ can be reproduced by the resulting pile-up that develops when $t$ is between $\sim 13 t_{\rm acc}$ and $\sim18 t_{\rm acc}$.  Such a non-steady state pile-up scenario has implications for the time evolution of the flare SED that could in principle be applied to the other SEDs reported in \cite{Buehler:2012}.  However, we do not pursue these implications here due to the other SED's higher contamination by the background nebula. Thus, a wide variety of electron distributions could have produced the observed SED shown in figure \ref{bestfit}, but we rule out standard shock acceleration scenarios with $p>2$ and suggest it was most likely a hard distribution of $p\lesssim 1$, possibly with a pile-up near the radiation reaction limit.

If the electron distribution is a hard power-law of $N\propto \gamma^{-1}$, where $N=0$ for $\gamma>\gamma_{\rm rad}$, the observer frame peak energy in the $\epsilon F_{\epsilon}$ representation is
\begin{align}
\epsilon_{\rm peak}'\approx0.6\hbar  \omega_c(\gamma_{\rm rad})\approx1.35 \eta\frac{\hbar m_ec^3}{e^2}\approx100 \eta\mbox{ MeV},
\label{max2}
\end{align}
where the critical synchrotron frequency is $\omega_c=3/2\gamma^2eB_{\perp}/m_ec$, and the location of the peak energy, $0.6\hbar\omega_c(\gamma_{\rm rad})$, was determined numerically.  The observed peak energy of a Doppler boosted SED is $\epsilon_{\rm peak}=\delta \epsilon_{\rm peak}'$, implying that any observed peak energy above 100 MeV must have a minimum Doppler factor of $\delta_{\rm min}= \epsilon_{\rm peak}/\epsilon_{\rm peak}'$.  Thus, for the April 2011 flare, $\delta_{\rm min}\gtrsim 375/100\sim4$.  

However, if the theoretical criteria discussed above are met for the electron distribution to become mono-energetic, then the SED will be effectively described by the single-particle synchrotron SED, whose form in the low energy limit is $\epsilon F_{\epsilon}\propto \epsilon^{4/3}$, and in the high energy limit, $\epsilon F_{\epsilon}\propto \epsilon^{3/2}\exp{(-\epsilon/\epsilon_c})$.  Since the $\epsilon F_{\epsilon}$ mono-energetic synchrotron spectrum peaks at the angular frequency of $\omega_{\rm peak}\approx 1.3\omega_c$, this implies the $\epsilon F_{\epsilon}$ peak from emitting particles with Lorentz factors of $\gamma_{\rm rad}$ is located at a photon energy of
\begin{align}
\epsilon_{\rm peak}'\approx1.3\hbar  \omega_c(\gamma_{\rm rad})\approx2.9 \eta\frac{\hbar m_ec^3}{e^2}\approx200 \eta\mbox{ MeV}.
\label{max}
\end{align}
Thus, the April 2011 flare implies a minimum Doppler factor of $\delta_{\rm min}\gtrsim2$.  Note that this is a lower estimate of $\delta_{\rm min}$ compared to the estimate made assuming a hard power-law electron distribution.  

\begin{figure}
\centering
\includegraphics[width=3.35in]{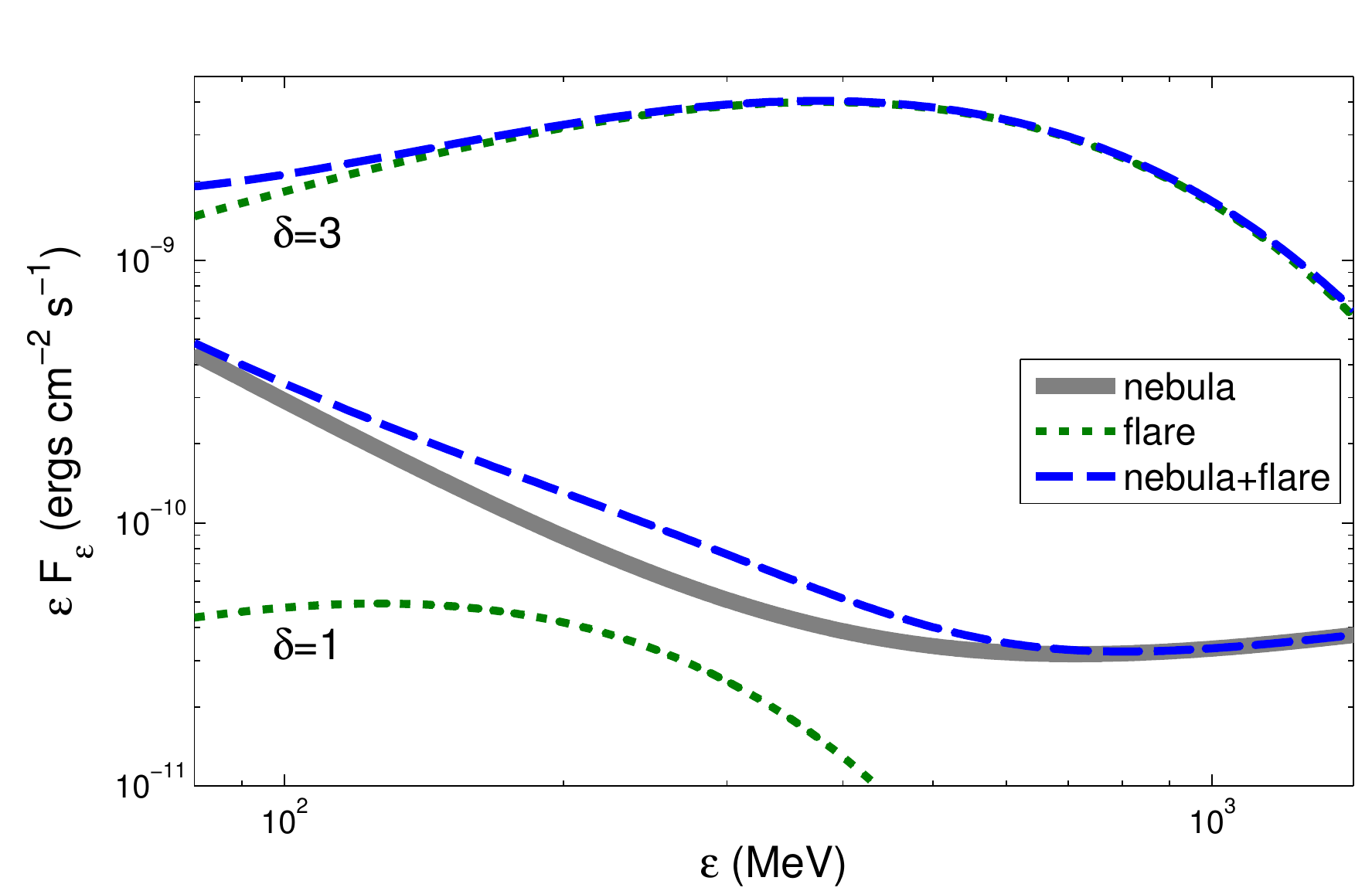}
\caption{Shown are Doppler boosted single-particle synchrotron SEDs (dotted lines) for $\delta=1$ and $\delta= 3$.  The normalization of the intrinsic SED and the critical frequency values are fixed so that the $\delta=3$ SED displays the maximum reported $\epsilon F_{\epsilon}$ value and correct peak energy during the April 2011 flare: $(\epsilon F_{\epsilon})_{\rm max}\sim 4\times 10^{-9}$ ergs cm$^{-2}$ s$^{-1}$ and $\epsilon_{\rm peak}=375$ MeV  \protect\citep{Buehler:2012}.  The average nebula SED (thick line) reported in \protect\cite{Buehler:2012} is summed with the flare SEDs to produce a combined SED shown by the dashed line.  Note that the difference between $\delta=1$ and $\delta=3$ SEDs is enough to render the former nearly unobservable compared to the average nebula emission.}  
\label{spectrum} 
\end{figure}

To illustrate the importance of Doppler beaming, we have plotted in figure \ref{spectrum} the same intrinsic single-particle synchrotron SED with two different Doppler factors, $\delta=1$ and $\delta= 3$.  The different Doppler factors affect the intrinsic SED's photon energies, $\epsilon=\delta \epsilon'$, and normalization, $\epsilon F_{\epsilon}(\epsilon)=\delta^4 \epsilon'F'_{\epsilon'}(\epsilon')$ \citep{Lind:1985}.  We adjust the normalization of the intrinsic SED to ensure that the $\delta=3$ flare has a maximum at $\epsilon F_{\epsilon}\sim 4\times10^{-9}$ ergs cm$^{-2}$ s$^{-1}$, as observed.  The same intrinsic normalization is used in the $\delta=1$ flare.  The constant Crab Nebula SED as described in \cite{Buehler:2012} is plotted as well.  Note that by using Doppler factors consistent with the SED cut-off observed during the most luminous part of the April 2011 flare ($\delta \gtrsim$ few), the unbeamed flare does not significantly increase the high energy flux of the average nebula. 

The non-detection of significant flaring by X-ray telescopes places a further constraint on the flare SEDs.  If the flare SED comes from a mono-energetic electron distribution, then in the \textit{Chandra} and \textit{XMM-Newton} energy bands, which we define here as extending from $\epsilon_{\rm min}=0.1$ keV to $\epsilon_{\rm max}=10$ keV, the SED goes as $\epsilon  F_{\epsilon}\propto \epsilon^{4/3}$.  For the single-particle synchrotron SED discussed above, where $(\epsilon F_{\epsilon})_{\rm max}\sim 4 \times 10^{-9}$ ergs cm$^{-2}$ s$^{-1}$ and $\epsilon_{\rm peak}=375$ MeV:
\begin{align}
F_X&=A\int^{\mbox{10 keV}}_{\mbox{0.1 keV}}{\epsilon^{1/3}d\epsilon}\approx 3.24\times 10^{-14}\mbox{ ergs cm$^{-2}$ s$^{-1}$} \notag \\
N_X&=B\int^{\mbox{10 keV}}_{\mbox{0.1 keV}}{\epsilon^{-2/3}d\epsilon}\approx 6.39\times 10^{-6}\mbox{ cm$^{-2}$ s$^{-1}$},
\label{xray}
\end{align}
where $F_X$ is the integrated X-ray flux and $N_X$ is the total photon flux.  These values are much smaller than the spatially integrated Crab Nebula X-ray energy flux of $\sim 10^{-7}$ ergs cm$^{-2}$ s$^{-1}$ and photon flux of $\sim 100$ cm$^{-2}$ s$^{-1}$ \citep{Kirsch:2005}.  Even the photon flux of the Crab nebula in a \textit{Chandra} resolution element of $\sim 1$ arcsec$^{2}$, which is $\sim 10^{-2}$ cm$^{-2}$ s$^{-1}$, is well above the photon flux we predict in the X-ray band.  If the flare emission were from a $p=1$ power-law electron distribution, then the estimates in equation (\ref{xray}) increase by a factor of $(\gamma_{\rm rad}/\gamma_{\rm min})^{2/3}$, assuming the lowest energy range of the SED goes as $F_{\epsilon}\propto \epsilon^{1/3}$.  Such a power-law could extend down to $\gamma_{\rm min}\sim 10^{-5}\gamma_{\rm rad}$ before the flare were comparable to the Crab Nebula flux in one \textit{Chandra} resolution element.  Hence, with a $\gamma_{\rm min}$ as low as $\sim 10^4$, it is possible to explain the non-detection of the flaring events by X-ray telescopes.

Our above discussion of the flaring SEDs has two implications: (i) for hard electron distributions, the MHD radiation reaction limit can lead to the formation of a pile-up electron distribution that is effectively mono-energetic, and (ii) significant emission beyond this limit implies the emitting region is moving along the line of sight at relativistic speeds.  Observations of the April 2011 flare suggest that, regarding (ii), a lower limit on the Doppler factor of $\delta \gtrsim$ few is required.  As for (i), the observed SED suggests a pile-up distribution that is not yet effectively mono-energetic.  Interestingly, as \cite{Buehler:2012} point out and we confirm, their data are not consistent with typical shock acceleration models, which usually produce $p\geq2$ \citep[e.g.][]{Kirk:2000}.  Instead, we suggest their observations are consistent with harder distributions ($p\lesssim1$) found in many magnetic reconnection models\footnote{Note that \cite{deGouveia:2005} construct a reconnection model that produces $p=2$ to 2.5 power-law indices, though \cite{Drury:2012} argues against their analysis.} \citep{Romanova:1992,Zenitani:2001, Zenitani:2007}.  Notably, some reconnection models even predict a $p=1$ electron distribution \citep{Larrabee:2003,Drury:2012}.  Thus, the April 2011 flare SEDs may be consistent with emitting particles accelerated in a reconnection region that are undergoing bulk relativistic motion with a Doppler factor of a few or more.

\section{A minijet statistical model}
\label{stats}
To illustrate how reconnection minijets relate to the high energy nebula flux and variability, we construct a toy model that produces statistical predictions about the high energy nebula light curve.  We postulate that reconnection minijets are random independent events in the nebula with an associated average reconnection event rate, $n_r$, and are therefore described by Poisson statistics.  A significant simplifying assumption we make is in presuming that the probability density functions (PDFs) for the intrinsic reconnection emission region timescale, $\tau'$, unbeamed intrinsic flux, $f'$, and Lorentz factor, $\Gamma$, are narrow enough to be treated as Dirac delta probability densities.  Thus, because $\tau'$, $f'$, and $\Gamma$ are constants, the statistics of the random variables $\tau$ (observed timescale) and $f$ (observed flux), are determined by the minijet Doppler factor, $\delta$, itself a random variable.  Another significant simplifying assumption we make is that the reconnection outflows have an isotropic angular distribution.  

The following discussion divides into two sections: \S\ref{fstats} covers the statistics of individual minijets, and \S\ref{tstats} develops time series statistics relevant to the nebular light curve as a whole. 

\subsection{Individual minijet statistics}
\label{fstats}

Define a spherical coordinate system $(r,\phi,\theta)$ with the z-axis ($\theta=0$) pointing along the line of sight so that the viewing angle, $\theta$, of any given jet is equal to the coordinate $\theta$ associated with its trajectory.  Thus, the PDF $\rho(\delta)$ is a function of the random variable $\theta$ alone.  Assuming the minijet angular distribution is isotropic so that $\rho(\delta(\theta))d\delta=d(\cos{\theta})$, then from the definition of the Doppler factor, 
\begin{align}
d(\cos{\theta})=\frac{1}{\beta}d\left(1-\frac{1}{\Gamma \delta}\right)=\frac{1}{\beta \Gamma \delta^2}d\delta.
\end{align}
Therefore, the Doppler factor PDF is
\begin{align}
\rho(\delta)d\delta=\frac{1}{\beta \Gamma \delta^2}d\delta,
\label{Pd}
\end{align}
where $\delta_{\rm min}=(\Gamma)^{-1}$ and $\delta_{\rm max}= (\Gamma-\sqrt{(\Gamma^2-1)})^{-1}$.  We note that $\delta_{\rm min}$ may effectively take on higher values, changing the normalization as well, so different values for cases where the debeamed flares gamma-ray flux is below a telescope's .

Using equation (\ref{Pd}) we now calculate observable quantities such as minijet timescale and flux distributions.  Regarding timescales, we find $n(\tau)d\tau$, which is the number of minijets that activate per unit time whose observed duration is between $\tau$ and $\tau+d\tau$.  This can be calculated by substituting $\delta=\tau'/\tau$ into equation (\ref{Pd}) and including a factor of $n_r$:
\begin{align}
n(\tau)d\tau=\frac{n_rd\tau}{\beta\Gamma\tau'} \quad \mbox{for $\frac{\tau'}{(1+\beta)\Gamma}<\tau<\Gamma\tau'$}
\label{ntau}
\end{align}
Hence, there is equal probability that short or long duration minijets will be observed.  To obtain the flux distribution, we assume the intrinsic gamma-ray flux of each mini-jet, $f'$ is Doppler boosted such that $f=\delta^q f'$, where $q=3+\alpha$ ($F_{\epsilon}\propto \epsilon^{-\alpha}$) for moving components, and for bolometric flux, $q=4$ \citep{Lind:1985,Jester:2008}.  We now substitute the Doppler boosting formula into equation (\ref{Pd}) to find the minijet flux probability density
\begin{align}
\rho(f)df=\frac{1}{q\beta\Gamma f'}\left(\frac{f}{f'}\right)^{-\frac{q+1}{q}}df.
\label{rhof}
\end{align}
As expected, this formula is identical to that used in the calculation of luminosity functions for Doppler beamed sources \citep[e.g. eqn. 2 of][]{Urry:1984}.  This equation implies that $\rho(f)\propto f^{-1}$ for $q \gg 1$, which means the minijet flux average, averaged over different flares, \textit{is dominated by rare bright flares}.

\begin{figure}
	\centering
		\includegraphics[width=2.3in, angle=270]{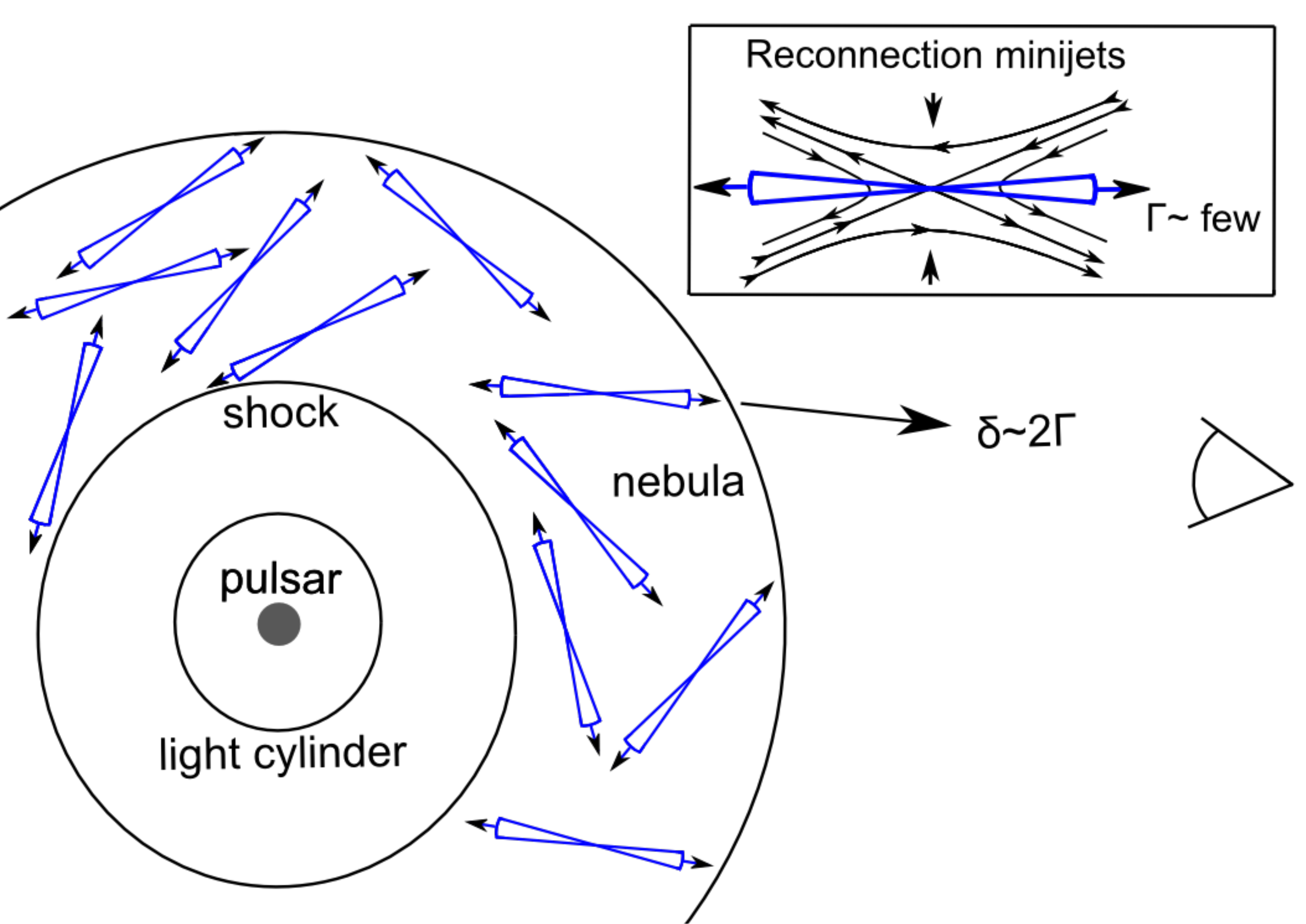}
		\caption{Cartoon schematic of reconnection sites in the nebula as viewed from above the toroidal plane (defined by the pulsar spin axis).  As shown in the upper right corner box, each reconnection site consists of plasma inflows into the reconnecting region and twin relativistic outflows (``minijets") with some Lorentz factor $\Gamma$ of order few.  Also shown in the schematic is a minijet directed toward the observer, which causes a flare since its emission is more highly beamed compared to its off-axis counterparts.}
	\label{fig:cartoon}
\end{figure}

\subsection{Minijet time series statistics}
\label{tstats}
In this subsection we derive quantities relevant to the entire high energy nebula light curve.  To do this we analyze the statistics related to the random variable $F_{\rm neb}$, representing the total nebular high energy flux.  First, we derive the statistical behavior of the random variable $k$, or the number of flares active at any given moment, and then apply this to the analysis of $F_{\rm neb}$.  Finally, assuming all nebular variability is due to minijets, we derive the nebular light curve's power spectrum.

The statistical behavior of the light curve depends sensitively on $\lambda$, the average number of reconnection events that are active at any given time.  To derive $\lambda$, we first note that the differential flare rate, $dn/d\delta$, immediately follows from equation (\ref{Pd}) as $dn/d\delta=n_r\rho(\delta)$ (recall $n_r \equiv$ average nebular reconnection rate).  Thus, the differential overlap number is $d\lambda=\tau dn$ (recall $\tau$ is the observed flare duration), since $\tau dn$ is the differential number of flares with Doppler factor $\delta$ that are activate within one observed flare duration $\tau$.  To calculate the average overlap number over an interval of Doppler factors from $\delta_{\rm min}$ to $\delta_{\rm max}$, we integrate to obtain
\begin{align}
\lambda &= \int^{\delta_{\rm max}}_{\delta_{\rm min}}{\frac{n_r\tau'}{\beta \Gamma \delta^3}d\delta} \notag \\
&=\frac{n_r\tau'(\delta_{\rm max}^2-\delta_{\rm min}^2)}{2\beta\Gamma\delta_{\rm max}^2\delta_{\rm min}^2} \notag \\
&\approx \frac{n_r\tau'}{2\beta\Gamma\delta_{\rm min}^2} \quad (\mbox{for } \delta_{\rm max}\gg\delta_{\rm min})
\label{lambda}
\end{align}

Since flaring consists of a random process of unrelated events governed by an average overlap number $\lambda$ within the nebula, the probability that $k$ flares are active/overlapping at any given moment is governed by the Poisson distribution, $P(k)= \lambda^k e^{-\lambda}/k!$.  We have verified the applicability of the Poisson distribution and our calculation of $\lambda$ (eqn. \ref{lambda}) to our model via Monte Carlo simulations with a variety of parameters values.

The duty cycle of minijets (i.e. the fraction of time when the nebula contains one or more active minijet) can now be calculated.  The duty cycle of flares with $\delta_{\rm min}<\delta<\delta_{\rm max}$ is simply the probability that at a time $t$, one or more flares ($k\geq1$) are active: $f_{\rm duty}=P(k\geq1)=1-e^{-\lambda}$.  In the non-overlapping limit, $\lambda\ll1$, the duty cycle reduces to $f_{\rm duty}\approx \lambda$.

We may now analyze the statistics of $F_{\rm neb}$, the random variable representing the total high energy flux of the nebula.  To this end, we introduce the random variable for the total minijet flux due to reconnection outflows, $F_r$, and the constant nebula flux, $F_{\rm const}$, possibly originating from the pulsar wind termination shock; the random variables $F_{\rm neb}$ and $F_r$ are related by
\begin{align}
F_r=F_{\rm neb}-F_{\rm const}.
\end{align}
Note that if only one minijet were active at time $t_0$, then $F_r(t_0)=f$.  In order to derive the statistical behavior of the directly observable $F_{\rm neb}$, we analyze the PDF of $F_r$, $\rho(F_r)$, where $\rho(F_r)dF_r$ is the probability that, at any given time, the high energy flux due to reconnection minijets is between $F_r$ and $F_r+dF_r$.  Each minijet is presumed to have a square pulse profile.  Importantly, our analysis assumes that $\rho(F_r)$ and its parameters remain constant over time intervals well in excess of the light curve's autocorrelation time, $\tau_{\rm auto}$.  If this assumption holds, then the moments of $\rho(F_r)$ may be compared with the observed moments of the light curve extracted from data spanning time intervals larger than the observed $\tau_{\rm auto}$. 

The exact moment generating function of $\rho(F_r)$ can be found, which contains the same information as $\rho(F_r)$ (see appendix for its derivation).  While any moment of $\rho(F_r)$ may be calculated, we provide the first three moments here for $f'=1$:
\begin{align}
\left\langle F_r\right\rangle &=\frac{n_r\tau'}{(q-2)\beta\Gamma}\left(f_{\rm max}^{1-2/q}-f_{\rm min}^{1-2/q}\right) \notag \\
\frac{\sigma_{F_r}}{\left\langle F_r\right\rangle}&=\frac{q-2}{(2q-2)^{1/2}}\frac{(\beta\Gamma)^{1/2}\left(f_{\rm max}^{2-2/q}-f_{\rm min}^{2-2/q}\right)^{1/2}}{(n_r\tau')^{1/2}\left(f_{\rm max}^{1-2/q}-f_{\rm min}^{1-2/q}\right)} \notag \\
\gamma_1&=\frac{(2q-2)^{3/2}}{3q-2}\frac{(\beta \Gamma)^{1/2}\left(f_{\rm max}^{3-2/q}-f_{\rm min}^{3-2/q}\right)}{\left(n_r\tau'\right)^{1/2}\left(f_{\rm max}^{2-2/q}-f_{\rm min}^{2-2/q}\right)^{3/2}},
\label{moments}
\end{align}
Where $\sigma_{F_r}=\left(\left\langle F_r^2\right\rangle-\left\langle F_r\right\rangle^2\right)^{1/2}$ is the standard deviation and $\gamma_1=\sigma_{F_r}^{-3}\left(\left\langle F_r^3\right\rangle-3\left\langle F_r\right\rangle\left\langle F_r^2\right\rangle+2\left\langle F_r\right\rangle^3\right)$ is the skewness; the brackets $\left\langle \right\rangle$ represent a time average.  As a check against our analytical work here, we have spot checked equations (\ref{moments}) with Monte Carlo simulations using a variety of parameters and found them to be consistent with one another.  Note that comparing the above moments to an observed light curve's moments allows for one to test whether the light curve is generated by beamed minijets per our model. To better discuss the significance of equations (\ref{moments}), we assume $f_{\rm max}\gg f_{\rm min}$, $\Gamma\gg1$, and $f_{\rm max}\approx (2\Gamma)^q$, so that equations (\ref{moments}) can be approximated as
\begin{align}
\left\langle F_r\right\rangle &\approx\frac{2^{q-2}n_r\tau'\Gamma^{q-3}}{q-2} \label{ave2}\\
\frac{\sigma_{F_r}}{\left\langle F_r\right\rangle}&\approx \frac{(2q-4)\Gamma^{3/2}}{((2q-2)n_r\tau')^{1/2}}\\
\gamma_1&\approx\frac{2(2q-2)^{3/2}}{(3q-2)}\frac{\Gamma^{3/2}}{(n_r\tau')^{1/2}}.
\end{align}
The variation about the average as expressed by the relative standard deviation ($\sigma_{F_r}/\left\langle F_r\right\rangle$) can easily take on small or large values depending on how $n_r\tau'$ compares with $\Gamma^3$.  Because $\gamma_1$ depends on $\Gamma$, $n_r$, and $\tau'$ in the same way as $\sigma/\left\langle F_r\right\rangle$ in this approximation, their ratio depends only on the beaming index and is expected to be of order unity:
\begin{align}
\frac{\sigma_{F_r}}{\left\langle F_r\right\rangle\gamma_1}\approx\frac{(2q-4)(3q-2)}{2(2q-2)^2}\sim 1,
\label{ratio}
\end{align}
where this ratio for reasonable values of the beaming factor ($q\geq3$) ranges from $\sigma_{F_r}/(\left\langle F_r\right\rangle\gamma_1)\sim 0.44$ to $0.75$.

Another important statistical representation of the light curve is the power spectrum, $P(\nu)$, which measures its variability on different timescales.  We compute $P(\nu)$ for light curves composed of square pulses, exponential pulses (i.e. zero rise time and exponential decay), and Gaussian pulses, all with a timescale of $\tau=\tau'/\delta$.  Since individual pulses are generated by a stochastic process, the phase information of the Fourier transform is unimportant, implying the power spectra for an individual pulse, $P_I(\nu)$, can be used to generate the light curve power spectrum thusly:
\begin{align}
P(\nu)=\int^{\tau_{\rm max}}_{\tau_{\rm min}}{n(\tau)P_I(\nu)d\tau},
\label{Pnu}
\end{align}  
where $n(\tau)$ is found in equation (\ref{ntau}).  For the square pulse and the exponential pulse, the single pulse power spectra are $P_I^{\rm sq}\propto \sin{(\pi\nu\tau)^2}/\nu^2$ and $P_I^{\rm exp}\propto(1+4\pi^2\tau^2\nu^2)^{-1}$ respectively (do to excessive length we do not include the spectrum for Gaussian pulses).  For the total power spectrum, when $\nu\ll\tau_{\rm max}^{-1}\approx(2\Gamma\tau')^{-1}$, the power spectrum is constant (i.e. white noise) since the frequency dependence drops out of the integrand.  For $\nu\gg\tau_{\rm min}^{-1}\approx2\Gamma/\tau'$, only the power spectrum for square and exponential pulses goes as $P(\nu)\propto \nu^{-2}$ (for the square pulses, there is oscillatory behavior on top of the $\nu^{-2}$ decay).  This can be summarized as follows:
\begin{align}
P(\nu)=\begin{cases}
\mbox{const.} & \mbox{for $\nu\ll(2\Gamma\tau')^{-1}$} \\
\mbox{transition to $\nu^{-2}$} & \mbox{for $(2\Gamma\tau')^{-1}<\nu<2\Gamma/\tau'$} \\
\nu^{-2} & \mbox{for $\nu \gg2\Gamma/\tau'$} 
\end{cases}
\label{Pnu2}
\end{align}
Unlike the square and exponential pulses, Gaussian pulses do not display power-law, they drop off faster (see fig \ref{power}).  While we have found analytical expressions for $P(\nu)$ for some beaming factors (e.g. $q=4$), we do not reproduce them here due to their excessive length.  However, the solutions to equation (\ref{Pnu}) are plotted in figure \ref{power}.  As a check against our analytical work, we include the discrete Fourier transform of a Monte Carlo simulated light curve composed of square pulses, which clearly follows the corresponding calculated power spectrum.
\begin{figure}
\centering
\includegraphics[width=3.37in]{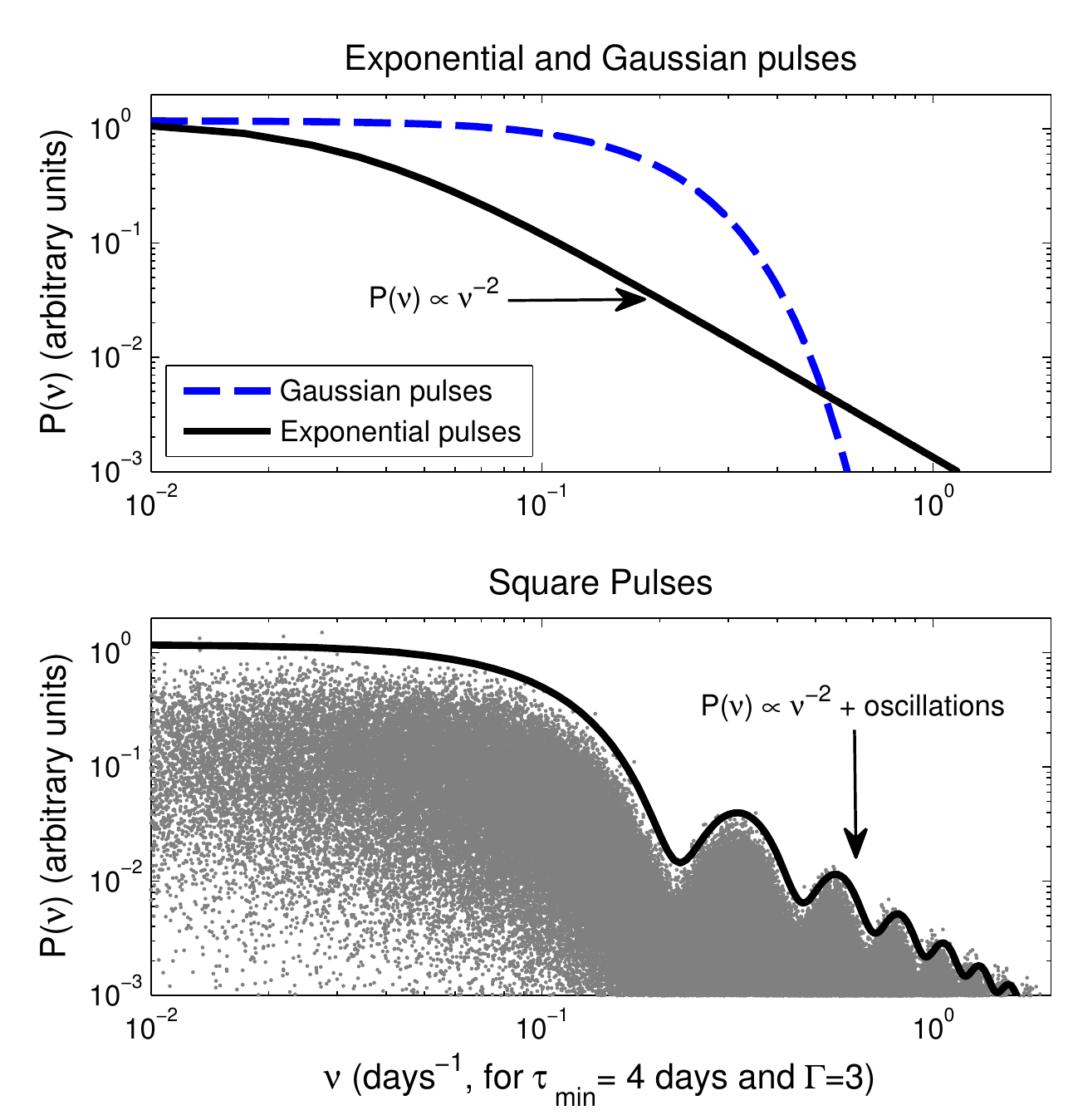}
\caption{Power spectra for exponential, Gaussian, and square pulses.  The exponential and Gaussian power spectra (top) display a marked difference, in that the exponential power spectrum decays as a power-lay ($\propto \nu^{-2}$) and the Gaussian power spectrum falls off more rapidly.  The square pulse power spectrum (bottom) is similar to the exponential one since it goes as $\nu^{-2}$, albeit with superimposed sinusoidal oscillations.   The gray data represents the square of the discrete Fourier transform (DFT) of the Monte Carlo generated light curve via square pulses.  The normalization of the square pulse $P(\nu)$ curve in the bottom panel is adjusted by hand for easy comparison with the simulated DFT (the exponential and Gaussian $P(\nu)$ curves use the same normalization).} \label{power}
\end{figure}

\subsection{Application of Statistical Model to Crab Nebula}

Our model may be divided into two categories: the non-overlapping regime and the overlapping regime.  In the non-overlapping regime, minijets do not temporally overlap and a significant constant emission component dominates, i.e. $\lambda\ll1$ and $F_{\rm const}/\left\langle F_r\right\rangle \gg 1$.  The opposite is true for the overlapping regime, where $\lambda\gg1$ and $F_{\rm const}/\left\langle F_r\right\rangle \lesssim 1$.  In the non-overlapping regime, each observed flare is due to an individual minijet's flux, $f$, hence individual minijet statistics are directly observable.  In this approximation  $F_{\rm const}/\left\langle F_r\right\rangle \gg 1$, so that $F_{\rm const}$ can be related to an observable quantity via $\left\langle F_{\rm neb}\right\rangle\approx F_{\rm const}$.  Therefore, we can write $f\approx F_{\rm flare}-\left\langle F_{\rm neb}\right\rangle$, where $F_{\rm flare}$ is the nebula flux ($F_{\rm neb}$) during an isolated flare. If we insert this expression into equation (\ref{rhof}) we obtain
\begin{align}
\rho(F_{\rm flare})dF_{\rm flare}\propto\left(F_{\rm flare}-\left\langle F_{\rm neb}\right\rangle\right)^{-\frac{q+1}{q}}dF_{\rm flare},
\end{align}
again, where $F_{\rm flare}= F_{\rm neb}$ when a single flare is active.  

A separate observable in this regime is the minijet timescale distribution (eqn. \ref{ntau}) which states that flares will be equally distributed across the allowed timescales.  

For the overlapping regime, the above analysis is complicated by the difficulty of associating a single flare with a single minijet.  However, applicable to both regimes is the light curve power spectrum (eqn. \ref{Pnu2}), which for some pulse profiles is $\propto \nu^{-2}$ in the short variability region.  However, this has been measured by \cite{Buehler:2012} as being $\propto \nu^{-1}$.  Such a measurement could be explained if the transition region where $P_{\nu}$ evolves from being constant to being $\propto \nu^{-2}$ is dominating that measurement. Other observables are the light curves moments (eqns. \ref{moments}), such as the ratio of variability to skewness (eqn. \ref{ratio}).  However, moments such as the relative standard deviation, which measures the light curve variability, cannot be directly compared with observational data since observed light curves smooth over short timescale variability because of time binning and the breaks in observational coverage.  
As an example light curve, figure \ref{lightcurve} represents of the non-overlapping regime of our model that qualitatively reproduces the Crab light curve.  The parameters used to generate figure \ref{lightcurve} are $(\delta_{\rm min},\delta_{\rm max},q,n_r,\tau',\Gamma,f')=(0.56,5.83,4,0.1875\mbox{ days}^{-1},24\mbox{ days},3,0.15\times10^{-7}\mbox{s}^{-1}\mbox{cm}^{-2})$, which imply a flare overlap number of $\lambda=1.25$.

\begin{figure}
\centering
\includegraphics[width=3.25in]{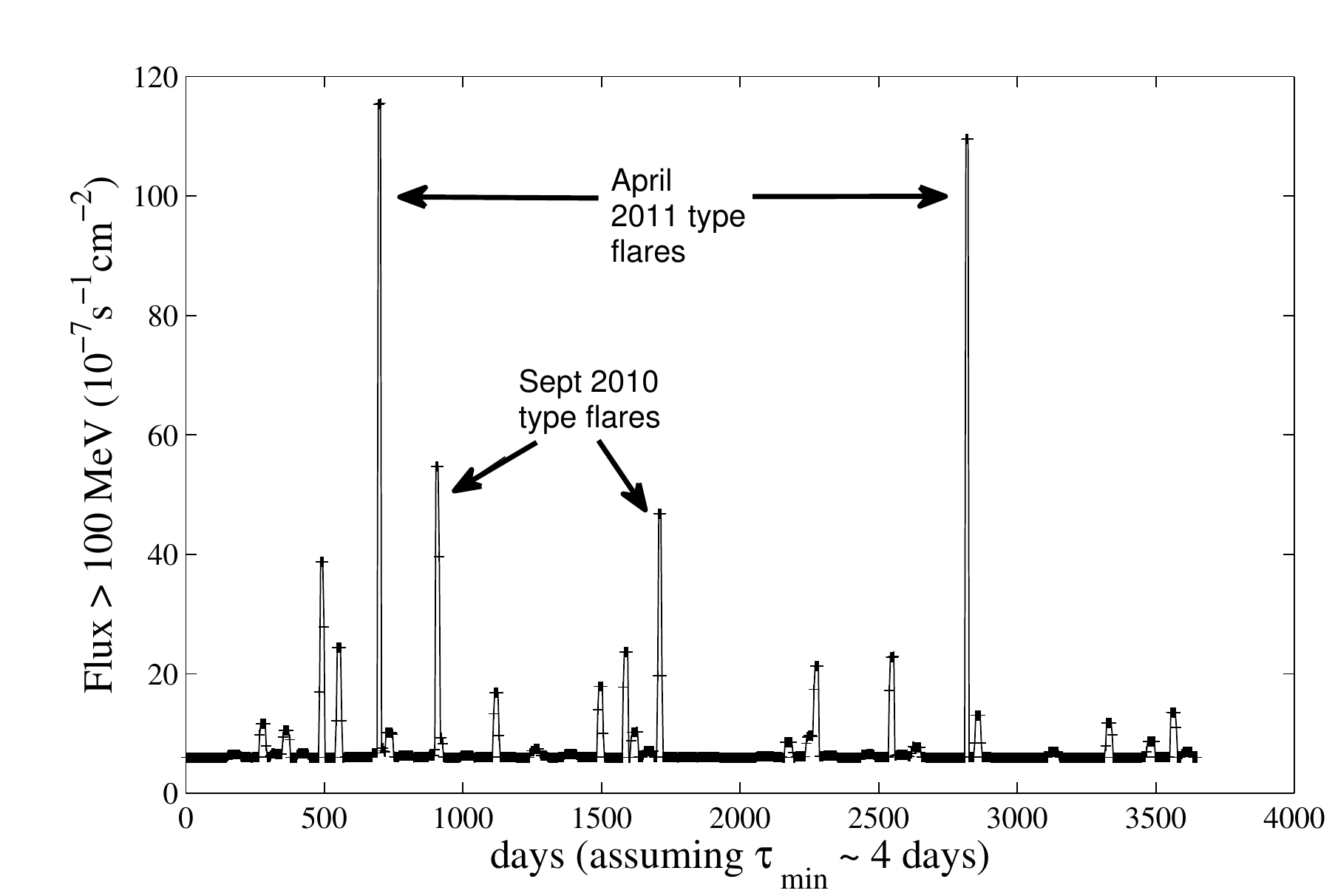}
\caption{Ten year simulated Crab nebula light curve.  The ``April 2011 type flares" represent flares with increases of $\sim$ 30 over the nebular average as found in \protect\cite{Buehler:2012}.  ``September type 2010" flares represent flux increases by $\sim5$ similar to the Crab September 2010 flare \protect\citep{Abdo:2011,Tavani:2011}.  The timescale assumes the shortest flare durations are 4 days, thus $\tau'\approx 2\Gamma \times 4$ days, so if $\Gamma=3$, then $\tau'=24$ days.  The simulated light curve was binned into 4 day intervals. For easy comparison to data, we add a steady component of $>$ 100 MeV flux $F_{\rm const}=6\times10^{-7}$s$^{-1}$cm$^{-2}$, consistent with the Crab Nebula's average flux as measured by Fermi/LAT \citep{Abdo:2011}.  The model parameters are $(\delta_{\rm min},\delta_{\rm max},q,n_r,\tau',\Gamma,f')=(0.56,5.83,4,0.1875\mbox{ days}^{-1},24\mbox{ days},3,0.15\times10^{-7}\mbox{s}^{-1}\mbox{cm}^{-2})$, with a corresponding flare overlap number of $\lambda=1.25$.} \label{lightcurve}
\end{figure}

For either regime, the energy in PWNe reconnection outflows must not exceed the total energy budget set by the spin-down power of the pulsar.  For example, if the non-overlapping model is true and approximately one flare per year is observed with a viewing angle of $\lesssim \Gamma^{-1}$, then $n_r\sim 4\Gamma^2$ yr$^{-1}$.  For the September 2010 flare parameters, this implies the minijet power expended is
\begin{align}
P_r=n_rE_r=4.0 \times 10^{35} L_{36.6} t_{5.5} n_r f_1^{-1}  \quad \mbox{erg s}^{-1},
\end{align}
which is much less than the Crab pulsar's spin-down power, $P_{\rm spin}\approx 5 \times 10^{38}$ erg s$^{-1}$.

In applying our statistical minijet model to the Crab Nebula, its simplifying assumptions that (a) the minijets' intrinsic parameters are the same and that (b) the minijet directions are isotropically distributed are both open to criticism.  Assumption (a) is challenged by the most luminous flare being longer in duration ($\sim$ 9 days) than the September 2010 flare ($\sim$ 4 days), since both assumption (a) and the Doppler transformations indicate the April 2011 flare should be shorter.  However, more flare observations are necessary before any firm conclusions are made regarding a correlation (or lack thereof) between observed flare luminosity and duration.  Regarding assumption (b), the clear toroidal morphology of the Crab nebula \citep[e.g.][]{Weisskopf:2000}, consistent with pulsar models with a toroidal outflow containing a toroidally dominated magnetic field and a large-scale current sheet, combine to make the assumption of an isotropic angular distribution of minijets questionable.  In the PWN split monopole model or the striped wind one \citep{Coroniti:1990,Bogovalov:1999}, reconnection in the implied current sheets would only produce minijets in the plane of the torus, rendering them unobservable since the pulsar spin axis is at an angle of $\sim 60^{\circ}$ to our line of sight \citep{Weisskopf:2000}.  For this reason, we have assumed the minijets are produced in an turbulent isotropic region of the nebula.  Nonetheless, the large-scale anisotropic morphology and structure of the magnetic field suggests that future work on this model should take into account at least some degree of anisotropy.

\section{Conclusions}
\label{conclusion}
We have constructed a statistical model of Crab Nebula high energy variability by assuming that magnetic reconnection sites throughout the nebula are activated randomly, and once activated, display emission characteristics controlled by the reconnection outflow Doppler factor.  At each site, a magnetically dominated reconnection region launches twin relativistic outflows along a randomly aligned axis.  GeV flares are observed when, by chance, a relativistic outflow is aligned with the line of sight and is thus Doppler-boosted.  The observed flares' unusually short durations and high luminosities suggest the emitting plasma is indeed moving toward Earth at relativistic speeds.

The flares' SEDs contain information about the particle acceleration process that suggests non-thermal particles are generated in reconnection regions, rather than by shocks, and are undergoing bulk relativistic motion along the line of sight.  Models of reconnection particle acceleration tend to create a hard power-law with an index close to $p\sim 1$ \citep{Romanova:1992,Zenitani:2001,Zenitani:2007,Larrabee:2003,Drury:2012}. We have shown that such distributions can easily form a pile-up near the radiation reaction limit implied by MHD, effectively becoming mono-energetic.  The April 2011 flare SEDs \citep{Buehler:2012} are inconsistent with shock acceleration and are instead consistent with a hard electron distribution that may contain a not yet effectively mono-energetic pile-up.  Furthermore, because the observed location of the SED peak is greater than the peak predicted by the synchrotron radiation reaction limit, the April 2011 flare's emitting region Doppler factor is $\gtrsim$ few.      

The predictions of our statistical minijet model can be summarized as follows:
\begin{itemize}
	\item When minijets do not temporally overlap one another, the probability density function for the nebula's high energy flux during a flare, $F_{\rm flare}$, is 
	\begin{align}
	\rho(F_{\rm flare})\sim \mbox{const}\left(F_{\rm flare}-\left\langle F_{\rm neb}\right\rangle\right)^{-1}.
	\end{align}
	\item The first three moments of the light curve may be compared with our theoretically calculated moments in equations (\ref{moments}), and any higher degree theoretical moments may be easily calculated using the method described in the appendix.
	\item The light curve power spectrum (eqn. \ref{Pnu2}) is constant (``white noise") for $\nu\ll (\Gamma\tau')^{-1}$, and goes as $P(\nu)\propto \nu^{-2}$ for $\nu\gg \Gamma/\tau$. 
\end{itemize}

Unlike the standard model for Crab Nebula non-thermal emission, wherein particles are accelerated by the pulsar wind termination shock \citep{Kennel:1984}, we have suggested here that magnetic reconnection may play an important or even dominant role in particle acceleration.  Further research will consider whether this reconnection model can explain both the steady nebula emission and the flaring and therefore preclude the need for shock emission altogether.  Our statistical model may also apply for AGNs that exhibit several minute TeV variability.

\section*{Acknowledgments}
We thank the Fermi/LAT team for making data concerning the April 2011 flare public and Konstantinos Gourgouliatos for many helpful discussions.

\onecolumn
\appendix
%
\section{Calculating the moments of $\rho(F_r)$}
\label{appendix}
Here we exploit a special property of moment generating functions \citep[MGFs, e.g.][]{Bulmer:1979} that allows us to calculate the MGF for the probability density function (PDF), $\rho(F_r)$.  First we state the expression for the probability that the observed minijet flux will be between $F_r$ and $F_r+dF_r$ at any given time:
\begin{align}
\rho(F_r)dF_r&=\sum^{\infty}_{k=0}{P(k)\rho(F_k)dF_r}=\tilde{\delta}(F_r)e^{-\lambda}dF_r+\sum^{\infty}_{k=1}\frac{\lambda^ke^{-\lambda}}{k!}\rho(F_k)dF_r,
\end{align}
where for compactness we have replaced $F_r|k$, or \textit{``$F_r$ given $k$"}, with $F_k$.  It is important to note that $\rho(F_r)$ is the PDF for the total summed minijet flux at any given timeslice, not the PDF that governs the flux of indiviudal flares, which is described by equation (\ref{rhof}).  

The MGF for a random variable $Y$ that follows the PDF $\rho(Y)$, is defined as $M_{Y}=\int^{\infty}_{-\infty}{\exp{(Yx)}\rho(Y)dY}$, which can be Taylor expanded to produce
\begin{align}
M_{Y}=1+\left\langle Y\right\rangle x+\frac{1}{2}\left\langle Y^2\right\rangle x^2+\ldots+\frac{1}{j!}\left\langle Y^j\right\rangle x^j,
\label{MGF1}
\end{align}
where $x$ is a dummy variable such that any moment of $\rho(Y)$ may be calculated via $\left\langle Y^j\right\rangle=d^jM_Y/dx^j|_{x=0}$.  Applying the definition of $M_{Y}$ to $\rho(F_r)$ we find
\begin{align}
M_{F_r}(x)&=\left\langle e^{F_rx} \right\rangle =e^{-\lambda}\sum^{\infty}_{k=1}{\frac{\lambda^k}{k!}\left\langle e^{F_rx} \right\rangle_k}=e^{-\lambda}\sum^{\infty}_{k=1}{\frac{\lambda^k}{k!}M_{F_k}},
\label{MGF2}
\end{align}
where $M_{F_k}$ is the MGF associated with conditional PDF, $\rho(F_k)$.

The special property of MGFs we use here is that for a random variable $F_k$ that is the sum of $k$ random variables, $F_k=\sum^{k}_{i=1}{f_i}$, the MGF is simply \citep{Bulmer:1979}
\begin{align}
M_{F_k}=\prod^k_{i=1}M_{f_i},
\end{align}
where $M_{f_i}$ is the moment generating function for the random variable $f_i$, the $i^{\rm th}$ minijet flux of the $k$ minijets active at time $t$ (the more debeamed minijet in each reconnection outflow pair is not taken into account as it contributes little to the overall flux).  Because the PDFs for all minijet random variables, $f_i$, are the same, then 
\begin{align}
M_{F_k}=M_{F_1}^k.
\label{MGFproperty}
\end{align}
$M_f$ is easily obtained from calculating the moments of $\rho(F_{k=1})$, which is equation (\ref{rhof}) with an extra factor of $\tau$ and a correspondingly different normalization.  This extra factor of $\tau$ is necessary because longer flares are more likely to be active at a time $t$ compared with shorter duration flares, while equation (\ref{rhof}) does not take a flare's time dependence into account.  We now calculate $\left\langle F_1^j\right\rangle$:
\begin{align}
\left\langle F_1^j \right\rangle &=\int^{f_{\rm max}}_{f_{\rm min}}{F_{k=1}^j\rho(F_{k=1}) dF_{k=1}}=A\int^{f_{\rm max}}_{f_{\rm min}}{f^j\rho(f)\tau df} \notag \\
&=\left(\frac{2}{jq-2}\right)\left(\frac{f_{\rm max}^jf_{\rm min}^{2/q}-f_{\rm max}^{2/q}f_{\rm min}^j}{f_{\rm max}^{2/q}-f_{\rm min}^{2/q}}\right),
\end{align}
where $f$ is the single flare flux and $A$ is a normalization constant.  Now $M_{F_1}$ can be obtained using equation (\ref{MGF1}), such that $M_{F_1}=\sum{1/j!\left\langle F_1^j \right\rangle}$, which can be inserted into equation (\ref{MGFproperty}) to produce
\begin{align}
M_{F_k}=M_{F_1}^k=\left(\sum^{\infty}_{j=0}{\frac{1}{j!}\left\langle F_1^j \right\rangle}\right)^k.
\end{align}
Hence, the total MGF associated with $\rho(F_r)$ is
\begin{align}
M_{F_r}=e^{-\lambda}\sum^{\infty}_{k=1}{\frac{\lambda^k}{k!}\left(\sum^{\infty}_{j=0}{\frac{1}{j!}\left\langle F_1^j \right\rangle}\right)^k}.
\end{align}
To find the first three moments of $M_{F_r}$, we calculate the first three moments of $M_{F_k}$ by expanding $M_{F_k}$ to third order in the dummy variable $x$:
\begin{align}
M_{F_k}&=M_{F_1}^k \notag \\
&\approx 1+k\left\langle F_1\right\rangle x+\left(\frac{k}{2}\left\langle F_1^2 \right\rangle+\frac{k(k-1)}{2}\left\langle F_1 \right\rangle^2\right)x^2+\left(\frac{k}{6}\left\langle F_1^3 \right\rangle +\frac{k(k-1)}{2}\left\langle F_1 \right\rangle \left\langle F_1^2 \right\rangle+\frac{k(k-1)(k-2)}{6}\left\langle F_1 \right\rangle^3\right) x^3
\label{Mfk}
\end{align}
Therefore, we can now read off the first three moments of the $\rho(F_k)$ distribution,
\begin{align}
\left\langle F_k \right\rangle &=k\left\langle F_1\right\rangle \notag \\
\left\langle F_k^2\right\rangle &=k\left\langle F_1^2\right\rangle +k(k-1)\left\langle F_1\right\rangle \notag \\
\left\langle F_k^3\right\rangle &=k\left\langle F_1^3\right\rangle+3k(k-1)\left\langle F_1\right\rangle\left\langle F_1^2\right\rangle+k(k-1)(k-2)\left\langle F_1\right\rangle
\end{align}
Now, note that
\begin{align}
\left\langle F_r^j\right\rangle=e^{-\lambda}\sum^{\infty}_{k=0}{\frac{\lambda^k}{k!}\left\langle F_k^j\right\rangle}.
\label{sums}
\end{align}
Evaluating the infinite sum in equation (\ref{sums}) for $k=1$, 2, and 3 leads us to the first three moments of the $\rho(F_r)$ distribution,
\begin{align}
\left\langle F_r \right\rangle&=\lambda\left\langle F_1\right\rangle \notag \\
\left\langle F_r^2\right\rangle&=\lambda\left\langle F_1^2\right\rangle +\lambda^2\left\langle F_1\right\rangle^2 \notag \\
\left\langle F_r^3\right\rangle&=\lambda\left\langle F_1^3\right\rangle+3\lambda^2\left\langle F_1\right\rangle\left\langle
F_1^2\right\rangle+\lambda^3\left\langle F_1\right\rangle^3.
\end{align}
Calculating higher order moments is straightforward.  In general, the $j^{\rm th}$ moment of $\rho(F_r)$ can be calculated by Taylor expanding the expressions for $M_{F_k}$ (eqn. \ref{Mfk}), reading off the $j^{\rm th}$ order term in said expansion, and using this term in evaluating the infinite sum for $\left\langle F_r^j\right\rangle$ (eqn. \ref{sums}).

\label{lastpage}
\end{document}